# Early Preconfiguration Failure：A Novel Predictor of the Repetitive Subconcussion

Jiajia Li [1,2], Zhenzhen Yu [1], Zhenghao Fu [2,3,4], Guozheng Xu [2,3,*], Jian Song [2,3,*]

[1]College of Information and Control Engineering, Xi'an University of Architecture and Technology, Shaanxi, Xi'an, 710055, China.

[2]Department of Neurosurgery, General Hospital of Central Theater Command, Wuhan, 430070, China.

[3]The First School of Clinical Medicine, Southern Medical University, Guangzhou 510515, China

[4]Department of Neurosurgery, Guangdong Sanjiu Brain Hospital, Guangzhou 510510, China.

*Corresponding authors at: Department of Neurosurgery, General Hospital of Central Theater Command, 627 Wuluo Road, Wuhan 430070, China.

**Email addresses:** songjian0505@smu.edu.cn (J. Song), xu-gz@163.com (G. Xu).

**This PDF file includes:**

    Main Text
    Figures 1 to 6
    Table 1

**Summary**

Early diagnosis and assessment of repetitive subconcussive (rSC) brain injuries are crucial for early clinical intervention. Conventional methods, largely relying on slow fMRI, fail to capture millisecond-level early cortical dynamics, particularly spatiotemporal features associated with pre-configuration dynamics. This study introduces a novel approach integrating dynamic hierarchical spatial features and cortical early behavioral time-domain sensitivity, utilizing EEG and visual attention tasks. We analyzed cortical early behaviors in 24 healthy controls (HC), 21 rSC patients, and a validation cohort of 25 cTBI patients from public datasets. Results reveal distinct temporal patterns in HC: elevated integration at 0–100 ms, rebound dynamics at 100–200ms, and visual perception integration peaks at 200–600 ms. In contrast, rSC patients exhibited significantly impaired dynamic features, with reduced integration levels indicating a decline in pre-configuration dynamics. Signed center distance (SCD) analysis of separation-integration trajectories showed significantly lower early SCD values in rSC patients compared to HC, while cTBI patients displayed negative SCD values, reflecting irreversible damage. Machine learning classification achieved optimal performance in distinguishing between HC, rSC, and cTBI groups using early cortical features, highlighting the critical role of millisecond-level cortical dynamics in rSC diagnosis.

**Keywords:** early cortical activity, pre-configuration, separation and integration, repetitive subconcussion

## 1. Introduction

The early detection of brain diseases—particularly those refractory to treatment—is critical for facilitating timely intervention and slowing disease progression[1]. Among these conditions, repetitive subconcussive (rSC) injuries stand out for their high prevalence[2], arising from diverse mechanisms including sports-related impacts, military exposures, high-intensity daily activities, and non-contact physical disturbances[2]. Pathologically, subconcussion (SC) injuries primarily manifest as axonal damage in cerebral neurons and compromised blood-brain barrier integrity[3,4]. However, their mild or transient symptoms frequently lead to underdiagnosis and undertreatment, potentially progressing to chronic traumatic encephalopathy (CTE) and other neurodegenerative changes[5]. Epidemiological data indicate that approximately 26.4% of SC injury cases develop significant cognitive impairment[6], underscoring the urgent need for reliable predictive methodologies to identify asymptomatic or subtly symptomatic cases. Such tools would optimize therapeutic windows and prevent early disease progression.

Initial efforts to predict brain injury-related disorders have focused predominantly on explicit analytical methods for brain signals, including ERP-based steady-state evoked potentials[7-11] and functional connectivity matrices derived from fMRI data[12-14]. However, current detection algorithms—relying on robust characterizations of brain states—face significant limitations in identifying subtle variations, particularly in real-world settings where interindividual variability and external noise pose substantial challenges[15,16].

Notably, the brain's hierarchical structure exhibits heightened representational sensitivity[17,18], enabling precise characterization of its multi-level organizational complexity. Perturbations at any hierarchical level can trigger cascading effects across the entire system, making this framework ideal for modelling pathological states. To analyze these dynamics, novel separation-integration algorithms have been developed, integrating both static[19,20] and dynamic[21,22] feature extraction across multiple scales and temporal dimensions. Empirical studies validate their effectiveness in predicting various neurological disorders, most notably Alzheimer's disease[23,24].

Despite these advances, fMRI technology—constrained by limited temporal resolution—may miss critical time-domain features, compromising the accuracy of disease prediction models[25-27]. This limitation is particularly impactful for bottom-up cognitive processing, where millisecond-scale neural response features of early cortical activity (ECA) are readily overlooked in conventional time-domain analyses.

In the mechanistic interrogation of ECA states, generative brain model theory[28,29] furnishes a pivotal conceptual scaffold. This theoretical construct posits that bottom-up processing entails not merely deep feedforward computation but also an innately preconfigured neural self-organizational mechanism. Temporally, "early cortical dynamics" manifest during this self-organizational epoch, functioning as a critical preparatory substrate for subsequent perceptual activation—thereby entrenching ECA as an indispensable constituent of bottom-up processing architectures.

Traditional investigations into the temporal dynamics of bottom-up processing have focused on post-stimulus steady-state responses, particularly the P200 (~200 ms) and P300 (~300 ms) components in event-related potentials (ERPs)[30-32], as well as gamma oscillations within the 200–300 ms window revealed by time-frequency analysis[33-36]. However, our recent study identified gamma oscillatory responses within the 100 ms window while investigating neural mechanisms of working memory[37]. This finding aligns with the dynamic neural activity patterns proposed by generative brain theory, highlighting the potential significance of early cortical responses (~100 ms or earlier) in neural information processing. Compared to the more pronounced response characteristics observed in later phases (200–600 ms), these early cortical dynamics may exhibit higher sensitivity and predictive validity for brain disease detection. Thus, deeper exploration of the neurophysiological mechanisms underlying ECA could yield more precise and sensitive biomarkers for the early diagnosis and risk assessment of rSC, with important implications for clinical translation.

To address these challenges, this study presents a novel analytical framework that integrates spatially sensitive hierarchical features with temporally precise attributes. Leveraging EEG responses to visual stimuli—captured with millisecond precision—we aim to develop predictive methodologies for identifying subconcussive impacts and assessing related injury outcomes.

We enrolled 24 healthy controls (HC) and 21 patients with rSC, employing a standardized GO-NOGO visual task (see Materials and Methods for details). Under identical experimental conditions, EEG responses were recorded to compare neural mechanisms between groups. Additionally, EEG data from 25 patients with chronic traumatic brain injury (cTBI) were collected under GO-NOGO perceptual stimulation for comparative analysis, aiming to validate our theoretical model's ability to distinguish attentional characteristics across traumatic brain injury (TBI) subtypes. As a mild TBI subtype[38](Figure 1A), rSC often presents with subtle clinical manifestations; behavioral analyses demonstrated no statistically significant disparities in accuracy (ACC) or reaction time (RTT) between the HC and rSC groups (Figure 1C). This lack of overt behavioral divergence necessitated an exploration of latent neural mechanisms, which was pursued through examinations of the segregation-integration dynamics of brain networks derived from ECA signals (Figure 1D).

Specifically, we systematically interrogated the dynamic properties of brain network segregation and integration during early (0–200 ms) and late (200–600 ms) post-stimulus epochs, with a focus on their divergent characteristics under healthy versus pathological conditions. We further evaluated whether early-phase (<200 ms) connectivity patterns in the HC group subserve perceptual pre-configuration for subsequent phases, and comparatively assessed perturbations of this pre-configurational capacity in the rSC group. By integrating the dynamic equilibrium between segregation and integration, we analyzed phase space distribution patterns across HC, rSC, and cTBI groups from a dynamical systems framework. Finally, four distinct machine learning approaches were employed for comparative validation to ascertain whether early-phase features confer enhanced predictive utility for disease diagnosis.

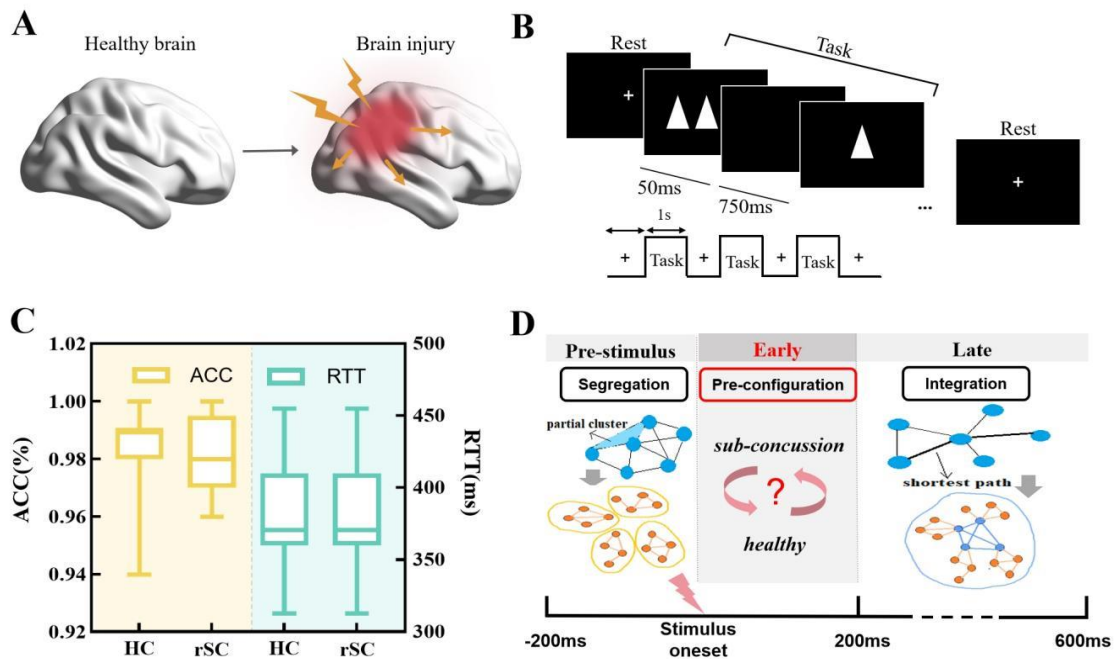

**Figure 1.** Concept-Experiment Framework rSC and Behavioral Analysis. (A) Schematic diagram illustrating brain-state transition from a healthy state to brain injury induced by repetitive subconcussive impacts. (B) The GO-NOGO visual task paradigm employs a double triangle as the target stimulus (Go) and a single triangle as the non-target stimulus (No-go). Each stimulus is presented for 50 ms, with inter-stimulus intervals randomized between 700–800 ms (mean: 750 ms). (C) Behavioral performance metrics for HCs and repetitive subconcussion individuals (rSCs): ACC, accuracy; RTT, reaction time. (D) Our proposed framework suggests that the critical early hierarchy structure during the pre-configuration stage may account for the transition between HC and rSC states. The dynamical brain network states are categorized into pre-stimulus (baseline), early, and late phases.

## 2. Methods

### 2.1. Participants and Stimuli

#### 2.1.1. Experimental data

This study enrolled 45 participants, comprising 24 healthy individuals recruited from college campuses and community settings, and 21 male parachutists with a history of repetitive concussive impacts. Participant inclusion was guided by previous epidemiological studies indicating that mild traumatic brain injury (mTBI) predominantly affects male adults aged 18–65 years[39], with the highest risk observed in young adults aged 18–23 years[40]. Accordingly, all participants in the current study were males aged 18–25 years. Groups were matched for age and years of education, with all participants being right-handed, Chinese-speaking, and having normal or corrected-to-normal vision with no color blindness. Cumulative exposure to indirect head impacts was quantified using the total number of actual and simulated platform jumps. Parachuters were included if they had completed at least 70 actual parachute jumps and 1,500 simulated platform jumps. Behavioral and EEG data were collected from parachuters within 1–3 weeks (10.18±3.5 days) of their last training session. Exclusion criteria were as follows: moderate-to-severe traumatic brain injury (TBI); history of craniotomy; neurological or psychiatric disorders; alcohol or substance abuse; and other conditions known to impair cognitive function. Additionally, HC participants were excluded if they engaged in organized sports to mitigate potential ceiling effects related to physical activity.

The Mini-Mental State Examination (MMSE) was administered to assess general cognitive function in all participants. Given that significant post-concussion symptoms can impact cognitive performance[41], the Post-Concussion Scale (PCS)[42] was used to evaluate symptoms in the rSC group; parachuters with PCS scores >5 were excluded. Written informed consent was obtained from all participants prior to experimentation. This prospective study was approved by the Ethics Committee of the General Hospital of Central Theater Command (approval no. [2020] 041–1).

A GO-NOGO perceptual stimulus task was performed using E-prime 2.0 software (Psychology Software Tools, Inc., Sharpsburg, PA). Participants sat in a quiet testing room, with their eyes positioned 70 cm from the screen and one hand placed near the keyboard for quick responses. Before the formal Electroencephalographic (EEG) recording, participants were informed of the experimental procedures and rules, and they completed a full practice session. During the experiment, there were two types of stimuli: a double triangle, which served as the target stimulus (GO), and a single triangle, which served as the non-target stimulus (NOGO). These stimuli were displayed at a physical brightness of 120 cd/m² on the central computer screen. Each stimulus presentation lasted for 50 ms, with a random inter-stimulus interval between 700 and 800 milliseconds (ms), averaging 750 ms. When the GO stimulus appeared, participants were required to quickly press any key on the keyboard; when the NOGO stimulus appeared, participants were instructed not to respond. The experiment was divided into three blocks, each containing 60 GO trials and 40 NOGO trials (Figure 1B). In total, HC group completed an average of 291.792 ± 9.198 trials, while rSC group completed 265.286 ± 31.379 trials (see Table S1 for individual participant data). Throughout the experiment, participants were required to fixate on the center of the screen, minimizing blinking and body movements to ensure concentration.

### 2.1.2. The OpenNEURO database (cTBI data)

Participants included 25 patients with cTBI (11 males, 14 females; age range 18–55 years), whose electroencephalogram (EEG) data were collected during a GO-NOGO perceptual stimulus task[43].

### 2.2. EEG recording and pre-processing

Electroencephalographic (EEG) signals were recorded using a 64-channel EEG system (Germany eego™ amplifier) with a sampling rate of 1000 Hz. The electrodes were positioned according to the international 10–20 system. 62 electrodes were used in this experiment, of which the FT9 and FT10 electrodes were not used and the online reference electrode was placed on the earlobe. The online reference electrode was placed on the earlobe, and the impedance levels were maintained below 5 kΩ. Continuous EEG data were band-pass filtered between 0.05 and 200 Hz. Throughout the experiment, environmental noise was strictly controlled to ensure the accuracy and reliability of the data.

### 2.3. PLV Functional Brain Network

This study focused on the investigation of gamma waves (30-45 Hz), within which we found significant neural oscillation rhythms (Figure S1). To construct dynamic functional connectivity (dFC)[44,45], we calculated the phase locking value (PLV)[46,47] between any two EEG signal channels. PLV is a metric that smooths the non-stationary and nonlinear responses between neural signals by separating the phase and amplitude of the signals[48,49]. Specifically, PLV directly assesses the synchronized phase coherence within a given frequency band, unaffected by amplitude variations[50,51]. During the calculation process, we determined the length of the sliding window to capture the temporal dynamics of the signals. This method provides a more indicative time-frequency analysis of EEG signal synchrony, helping to reveal functional connections between brain regions under specific tasks or states. The definition of PLV in Eq. (1)[50]:

$$\mathbf{PLV}(t) = \left| \frac{1}{\delta} \int_{t-\delta/2}^{t+\delta/2} e^{j(\phi_y(\tau) - \phi_x(\tau))} d\tau \right| \tag{1}$$

Here, $\phi_x(\tau)$ and $\phi_y(\tau)$ represent two different signals, and the instantaneous non-envelope phase of the original signal at time point t is extracted through the Hilbert transform at a given frequency. $\delta$ denotes the width of the time window. The range of PLV values is from 0 to 1, with larger values indicating stronger synchrony between the calculated pair of channels, and smaller values indicating weaker synchrony. This study calculated the PLV for 62 pairs of channels based on this method, generating a series of time-resolved weighted undirected matrices for functional connectivity FC of size 62×62.

### 2.4. Nest-Spectral Partition (NSP)

The NSP method effectively quantifies multi-level functional segregation and integration based on feature modes[52]. The brain functional connectivity matrix FC at each time point is symmetric and represented as $\mathbf{FC} = \mathbf{X}\lambda\mathbf{X^T}$, where $\lambda$ represents the eigenvalues and $\mathbf{X}$ represents the eigenvectors. The NSP method detects brain functional modules at different levels by analyzing the eigenvectors.

At the first level, all channel eigenvector values are either positive or negative, and the entire brain network is considered a single module. At the second level, channels with positive eigenvectors are assigned to one module, while channels with negative eigenvectors form another module. At the third level, based on the signs of the eigenvectors within the modules from the second level, further subdivision into sub-modules occurs. This process progressively divides the functional connectivity network into multiple hierarchical modular structures.

At any given level, if all channel eigenvector signs within a module are consistent, that module cannot be further divided at the next level. This partitioning process terminates when only one channel remains in each module. After the partitioning is complete, the number of modules M and the module size m are obtained.

*2.5. Hierarchical Segregation and Integration*

Separation and integration have a complex relationship in terms of hierarchy. To reflect the connection between hierarchical separation and integration, the weighted number of modules at each layer can be defined in Eq. (2)[22].

$$H_i = \frac{\Lambda_i^2 M_i (1-p_i)}{N} \quad (2)$$

Among them, $\Lambda$ is the eigenvalue; $M_i$ represents the number of modules at the i[th] layer; N is the number of rows or columns in the FC matrix; $p_i = \sum |m_{ij} - 1/M_i|/N$ is a correction factor that reflects the deviation of the module size at this level from the size of uniformly distributed modules; $m_{ij}$ represents the size of the j[th] module at the i[th] layer.

At the first level of analysis, the functional connectivity (FC) network is viewed as a whole, with all nodes belonging to the same module. At this level, the entire network is not divided into smaller sub-modules. Therefore, using the network structure at this level, the global integration characteristics of the entire network can be calculated in Eq. (3)[22,50].

$$H_{in} = \frac{H_1}{N} = \frac{\Lambda_1^2 M_1 (1-p_1)}{N^2} \quad (3)$$

In the first layer, the entire network contains only one module, therefore $p_1=0$. Since no other modules exist, there is no need to make any corrections to the global integration components.

The separated components from the second layer to the n[th] layer provide a multi-level analysis of the network's separation structure. By calculating the average $H_i$ of these layers, the overall degree of separation in the brain's different hierarchical functional organization can be quantified in Eq. (4)[22,50].

$$H_{se} = \sum_{i=2}^{N} \frac{H_i}{N} = \sum_{i=2}^{N} \frac{\Lambda_i^2 M_i}{N^2}(1-p_i) \quad (4)$$

This study analyzed data under task conditions and calculated the $H_{in}$ and $H_{se}$ values for each of the 71 time windows, thereby capturing and quantifying the dynamic of functional separation and integration in the brain at different time points. These results provide a foundation for exploring the brain's instantaneous responses and dynamic adjustments in functional organization within 100 ms before and after stimulus presentation while performing specific tasks.

*2.6. Signed-Centroid Distance (SCD)*

The specific definition is shown in Eq. (5):

$$SCD = \begin{cases} \dfrac{|X_0 - Y_0|}{\sqrt{2}}, & Y_0 > X_0 \\ -\dfrac{|X_0 - Y_0|}{\sqrt{2}}, & Y_0 < X_0 \end{cases} \quad (5)$$

Where $X_0$ is the separation value of the centroid of the dataset, $Y_0$ is the integration value of the centroid of the dataset, and the centroid is the mean value of the dataset. This formula expresses the distance from the centroid of mass of the integrated-separated phase diagram to the balance ($H_{in} = H_{se}$). A positive sign indicates that the data point is located in the integrated advantage area ($H_{in} > H_{se}$), while a negative sign indicates that it is located in the separated advantage area ($H_{in} < H_{se}$).

## 3. Results

*3.1. Network Dynamics of Segregation-Integration during the ECA Phase in HCs and rSCs*

A large number of studies have shown that the brain network has a hierarchical modular organization structure[53-55]. This structural characteristic is considered an important basis for the brain's efficient execution of diverse cognitive tasks and is significant for studying the mechanisms of separation and integration within brain networks[22,52]. Using phase-locking value (PLV)[44,45], we constructed 62-channel gamma-band (30–45 Hz) dynamic functional networks (Figure S1) and quantified comparisons between HC and rSC groups during early and late post-stimulus phases by nested spectral partitioning (NSP)[50,52]. The analysis focused on integration ($H_{in}$), segregation ($H_{se}$), and integration-segregation difference (ISD) metrics (Figure 2A-C).

As shown in Figure 2, the HCs demonstrated a progressive enhancement of the whole-brain integration factor feature during the early phase (0–100 ms) following stimulus presentation, reaching a local peak (ECA peak) around 100 ms (Figure 2A). In contrast, the integration feature in the rSC group did not exhibit a significant increase during the 0–100 ms period, and only a non-significant ECA peak was observed. This difference was further confirmed in the early-phase statistical results in Figure 2D: the degree of whole-brain integration enhancement in the rSC group was significantly lower than that in the HC group ($p < 0.01$). According to the integration-segregation balance theory[21], the integration and segregation features are negatively correlated: when integration strength increases, segregation strength decreases correspondingly, and the integration-segregation difference (ISD) between the two increases simultaneously. Using ISD = 0.5 as a reference line, Figure 2B and 2C show that the rSC group exhibited stronger segregation and lower ISD values in the early phase (Figure 2E and 2F, statistically significant $p < 0.001$). This result aligns with previous studies, which suggest that normal perceptual processing requires early whole-brain integration enhancement to support the pre-processing of external information[53]. The weakened ECA peak in the rSC group indicates impaired early pre-processing integration ability.

Furthermore, the HC group displayed typical rebound dynamics in the 100–200 ms phase: the integration decreased after the ECA peak and then increased again, maintaining a high integration steady state in the late phase (300–500 ms) (Figure 2A). This dynamic characteristic reflects the brain's sparse release process after early pre-configuration and the dynamic reintegration mechanism for subsequent perceptual tasks[56,57]. Comparing Figure S2A and S2C, it can be observed that the 300–500 ms period corresponds to the high-amplitude steady-state ERP response, marking the core phase of perceptual processing. This period overlaps highly with the sustained high integration factor duration of the late-phase rebound reintegration. However, it is noteworthy that the integration factor in the rSC group exhibited unique temporal characteristics: it decreased after 100 ms, then gradually increased, forming a local peak around 240 ms. Although the ERP amplitudes in the rSC group (Figure S2B and S2D) were slightly reduced compared to the HC group, the group still retained the complete signal pattern of the steady-state perceptual processing response in the 300–500 ms range. Additionally, the behavioral results (reaction time RTT and ACC) did not show significant differences between the rSC and HC groups (Figure 1C). This series of evidence indicates that the rSC group successfully completed the information processing of the perceptual stage.

Our subsequent analysis suggested that although the rSC group did not exhibit significant pre-configuration characteristics within the early 100 ms, it showed a delayed pre-configuration process between 200–300 ms. To define this process as pre-configuration, it is necessary to associate it with the memory retrieval mechanisms in the generative models of specific brain regions[22,53]. A detailed explanation of this will be provided in the results section.

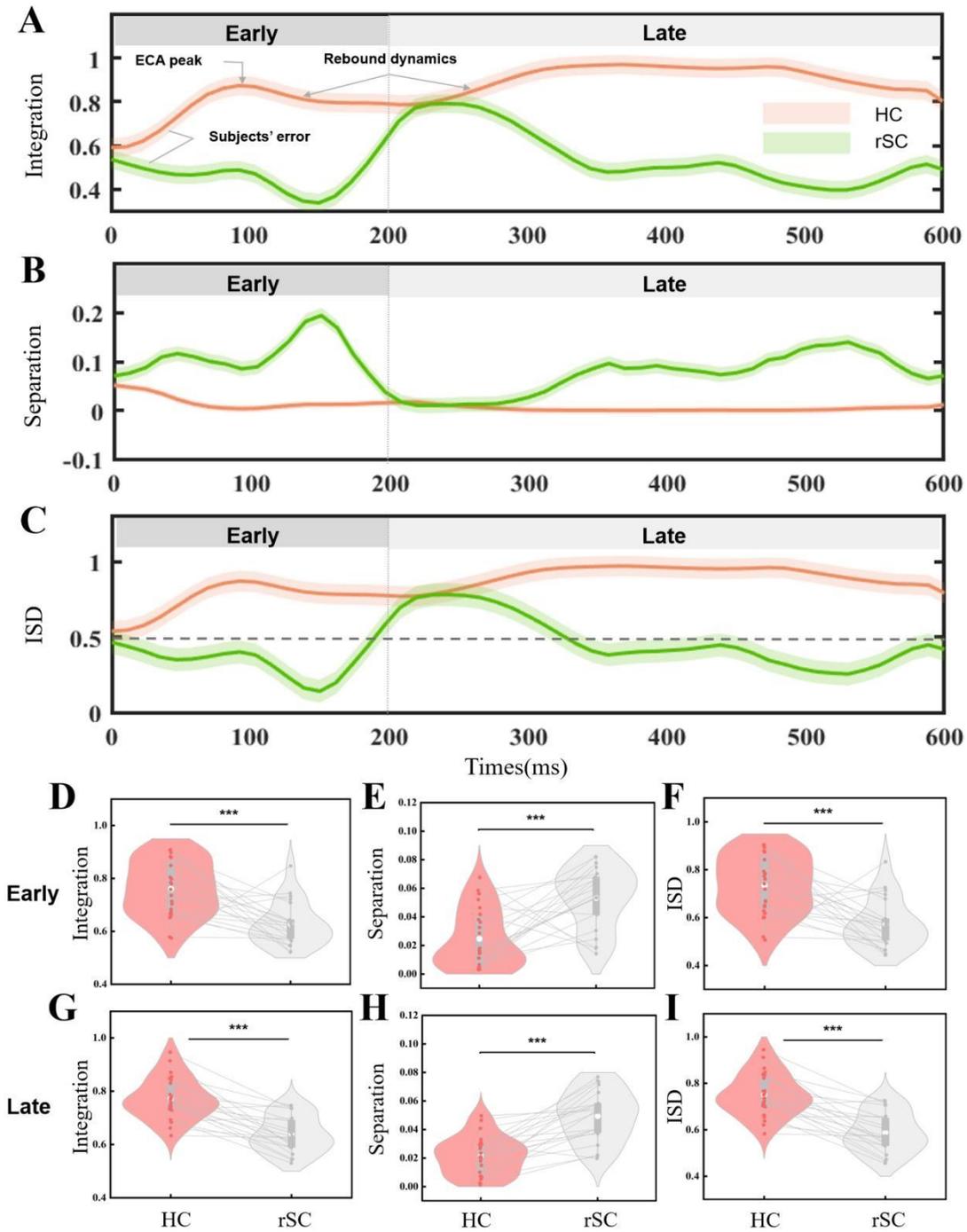

**Figure 2.** Early pre-configuration impairment in rSCs relative to HCs during dynamic integration, separation, and ISD processes.. (A-C) Mean values of dynamic integration (A), separation (B), and ISD (C), with shaded SEM areas (orange: HCs; green: rSCs) and corresponding trajectories (orange: HCs; green: rSCs). (D-I) Statistical comparisons of averaged integration (D, G), separation (E, H), and ISD (F, I) between HCs and rSCs. Significant differences were observed in both the early (D-F) and late (G-I) phases (all comparisons: $p < 0.001$). Analyses were conducted across subjects, regions, and channels.

*3.2. ROI Analysis of Pre-configuration Deficit in the ECA Phase*

To examine the relationship between the dynamics of pre-configuration and early integration, as well as their role as the "preparation stage" for late- phase perceptual processing, this study conducts a systematic analysis based on the theoretical framework of generative brain models[29]. The brain's processing of sensory information does not rely solely on a simple feedforward mode, but rather a parallel mode that integrates both feedforward and self-organized processes (corresponding to the right and left sides of Figure 3, respectively).

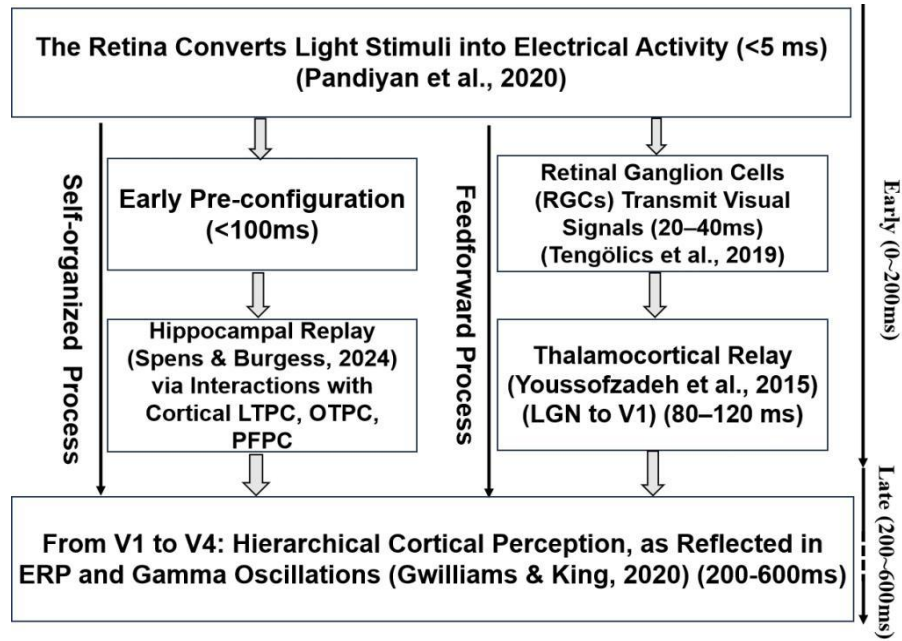

**Figure 3.** Schematic of Parallel Visual Processing Pathways, including a Self-Organized Brain Visual Processing Mechanism Driven by Early Pre-Configuration. This model illustrates a self-organized process for visual processing (left panel): During the early phase (0–200 ms), pre-configuration facilitates the retrieval of prior experiential memory information (hippocampal relay[58]) via interactions with intercortical cross-regional connections involving the left temporal-parietal cortex (LTPC), occipitotemporal-parietal cortex (OTPC), and prefrontoparietal cortex (PFPC)—all regions discussed in this paper ($p < 0.001$ for connection-related analyses). This retrieved memory information is associated with the incoming visual input, such that the framework does not depend solely on a feedforward process (right panel). In contrast, during the later phase (200–600 ms)—where ERPs P200 and P300 are detected(see Figure S2)—the steady-state response of cortical gamma oscillations further supports this visual processing[36,37,59,60].

The feedforward process includes: the retina converting light stimuli into electrical activity (<5 ms)[61]; retinal ganglion cells (RGCs) transmitting visual signals (20–40 ms)[62]; thalamocortical relay from the lateral geniculate nucleus (LGN) to the primary visual cortex (V1) (80–120 ms)[63]; and hierarchical cortical perception from V1 to V4, as reflected in ERPs and gamma oscillations (200–600 ms)[64] (see the right panel of Figure 3). For instance, the time-coding characteristics of ERPs in the 300–500 ms range (Figure S2A and S2C) exemplify this typical forward processing pattern (right panel of Figure 3).

By contrast, the self-organized process is primarily attributed to a critical early pre-configuration stage spanning 0–200 ms (left panel of Figure 3). During this stage, the memory system employs a rapid search-and-matching mechanism via hippocampal replay[58] (left panel of Figure 3) to retrieve memory-encoded information relevant to current perceptual input. It then applies a generative approach, referencing real-time input, to achieve rapid recognition of visual stimuli[29]. Meanwhile, according to previous findings, cortical cross-regional interactions during this memory replay process play significant roles—particularly involving the left temporal-parietal cortex (LTPC)[65], occipito-temporal-parietal cortex (OTPC)[66], and prefronto-parietal cortex (PFPC)[67]. The activities of these regions are key focuses of subsequent investigation, aimed at clarifying when and how the brain utilizes pre-configured networks to support memory retrieval functions.

To systematically investigate the differences in neural integration characteristics between the HC group and the retinitis pigmentosa with rSC group during the critical time window of perceptual processing, we propose the following hypothesis: During the 0–100 ms phase of stimulus processing, the HC group can establish a "pre-configured state" to prepare for phase memory retrieval. In contrast, rSC patients fail to complete this pre-configuration. Notably, the completion of visual perceptual cognition in rSC patients does not rely on fully intact neural activity patterns; instead, it is achieved through the delayed activation (220–270 ms) of relevant brain subregions. This delayed activation further triggers a compensatory, delayed hippocampal replay process, which ultimately supports information integration for visual perception. To test this hypothesis, we compared the gamma-band (30–45 Hz) functional connectivity networks of the two groups during two key time phases after stimulus presentation: 0–100 ms and 220–270 ms.

FC was visualized using BrainView software (https://www.brainview.com/), as shown in Figure 4. For consistency, a uniform threshold was applied for binarization, and only high-strength FC (strength > 0.85) was retained for visualization. In the figures, inter-regional connections are represented by blue lines; nodes (brain regions) are color-coded according to the following scheme: frontal cortex (red), temporal cortex (dark blue), central cortex (yellow), parietal cortex(green), and occipital cortex (light blue).

Our research will focus on integrating analyses of regions of interest (ROIs), with an emphasis on the significance of pre-configuration dynamic generative models and exploration of potential ROI damage patterns in individual rSC patients.

Previous studies have indicated that visual working memory representation primarily depends on functional coupling between the parietal and occipitotemporal cortices[66]. Additionally, functional impairments in the left temporal and parietal cortices have been identified as key factors contributing to visual information processing deficits following brain injury[65]—a finding closely linked to shape and spatial memory, both of which are relevant to the shape recognition task employed in this study[68]. In N-back tasks, functional connectivity between the prefrontal cortex (PFC) and parietal cortex has been shown to play a critical role in cognitive control and fluid intelligence[67]. These insights into the functions of relevant ROIs may provide foundational theoretical support for constructing the brain's generative model[69].

Building on this evidence, our analysis will focus on functional connectivity differences between the HC and rSC groups within the left temporal-parietal cortex (LTPC), occipitotemporal cortex (OTPC), and prefrontal-parietal cortex (PFPC).

As shown in Figure 4, the HC group exhibited significantly more high-intensity functional connections in the LTPC, OTPC, and PFPC, with these connections primarily distributed across multiple brain regions (Figure 4A and 4B). Notably, during the 0–100 ms pre-configuration phase, the HC group showed significantly enhanced connection strength in these three subregions, whereas the rSC group displayed a trend toward weakened connections (Figure 4C–E; statistical significance: $p < 0.001$).

In contrast, during the 220–270 ms time window, the rSC group exhibited delayed connection enhancement in the LTPC, OTPC, and PFPC. This finding indicates that the rSC group compensates for early neural integration deficits through late-phase compensatory neural activity—validating our hypothesis that even in states of subclinical brain injury, patients can complete visual cognition via delayed hippocampal replay processes.

Supporting this, Figure S2B and S2D shows that the rSC group still exhibited intact ERP activation peaks; additionally, their behavioral performance (ACC and RTT) did not differ significantly from that of healthy controls (see Figure 1C). These results confirm that the rSC group can still perform visual perception tasks normally, further demonstrating that their impairment is not severe: cognitive function remains largely intact, and information processing is instead achieved through a delayed neural integration mechanism.

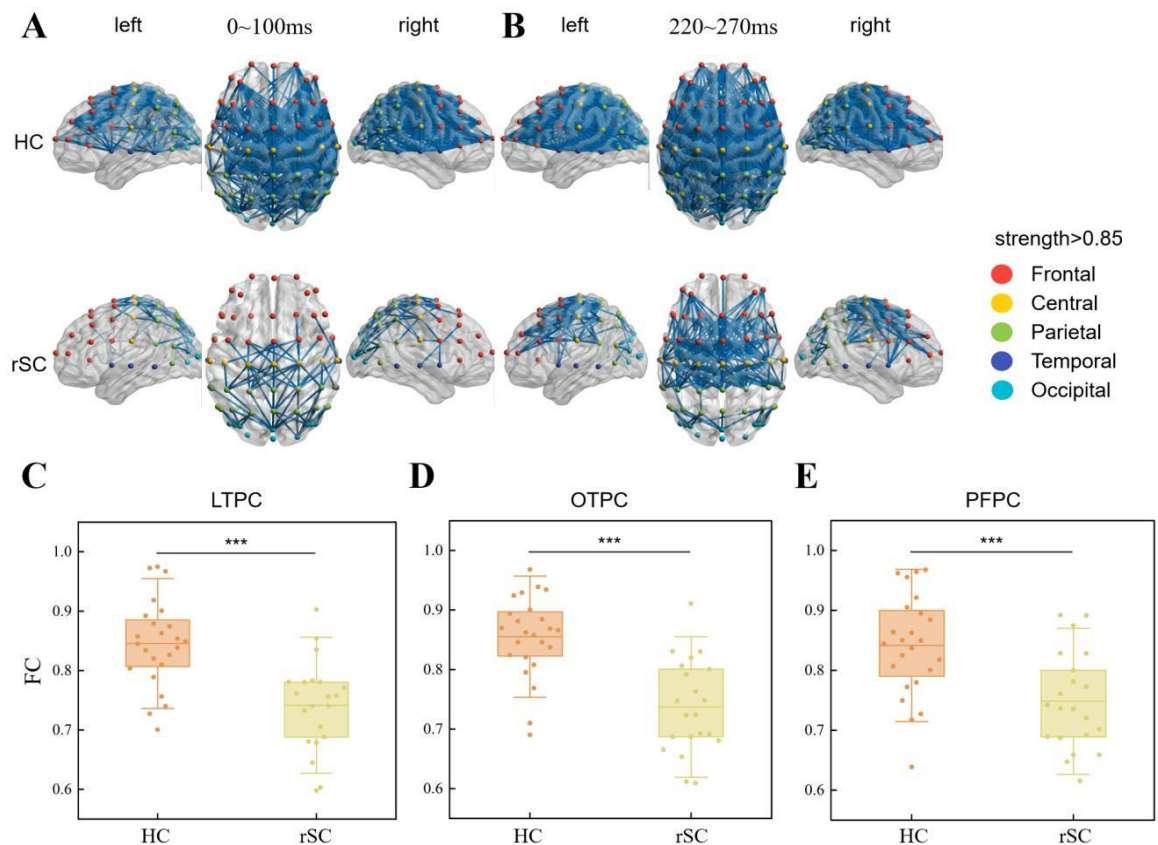

**Figure 4.** Brain Connectivity and Statistical Analyses Comparing HC and rSC Groups. (A) Comparison of high-strength connections (strength > 0.85) between HC and rSC groups during the 0–100 ms window. (B) Comparison of high-strength connections (strength > 0.85) between HC and rSC groups during the 220–270 ms window. Brain nodes are color-coded by lobe: frontal cortex (red), central cortex (yellow), parietal cortex (green), temporal cortex (dark blue), and occipital cortex (light blue). (C–E) Comparative results of functional connectivity (FC) densities in cross-regional connections involving the LTPC, OTPC, and PFPC rSCs and HCs groups. All statistical p-values were derived from paired t-tests comparing the rSCs and HCs groups, and all $p < 0.001$. Notably, the data used for the above statistical analysis were obtained from the early temporal phase (0–200 ms).

Table 1 systematically summarizes the overall degenerative trends and intergroup differences in network reintegration observed in the rSC group relative to the HC group, providing a more comprehensive understanding of the neural basis underlying this compensatory mechanism.

**Table 1.** Quantitative Analysis of Pre-Configuration Dynamic Characteristics.

|     | age   | sex | subjects | Hin/peak (0-100ms) | LTPC | OTPC | PFPC |
| --- | ----- | --- | -------- | ------------------ | ---- | ---- | ---- |
| HC  | 18-25 | M   | 24       | 0.875±0.015 | 0.845±0.073 | 0.855±0.068 | 0.842±0.085 |
| rSC | 18-25 | M   | 21       | ↓0.489±0.014 | ↓0.742±0.076 | ↓0.737±0.079 | ↓0.748±0.081 |

*3.3. Quantitative Analysis of cTBI, rSC, and HC Groups Using Early Integration-Separation SCD*

We have specifically analyzed the differences in the dynamic values of integration factors between rSC and HC groups during the early and late phases (Figure 2). To validate the generalizability of this differential pattern, we have extended our investigation to more severe cases of cTBI. As the Mayo Clinic Brain Bank reports that rSC can potentially lead to a transformation rate of up to 32% from early subconcussion to chronic traumatic encephalopathy[70]. The underlying changes related to this transformation should be address.

As shown in Figure 5A, as early concussion progresses to cTBI, the brain undergoes irreversible neuropathological changes. These include elevated levels of glial fibrillary acidic protein (GFAP) and neurofilament light (NfL), as well as tau protein accumulation[71-74]. Such changes may lead to a fundamental reconstruction of the integration factor.

To assess the generalizability of our findings, we selected EEG data from 25 participants with cTBI from the OpenNEURO database (n=25) (the cTBI dataset is from the OpenNEURO database shared by[43]) and conducted a comparative analysis with the HC and rSC group. We examined the dynamic distribution of integration factor phase trajectories in both early and late phases across all groups. Using Eq. (5), we quantitatively analyzed the distribution patterns of the Signed Centroid Distance (SCD) from the phase trajectory centers to the separation-integration balance line ($H_{in} = H_{se}$) for each group (Figure 5B and 5C).

The results demonstrated that in both early and late phases, the separation-integration trajectories of the HC and rSC groups were distributed above the balance line ($H_{in} > H_{se}$, shaded area), whereas those of the cTBI group were distributed below the balance line ($H_{in} < H_{se}$, area below the shaded region) (Figure 5B).

These findings suggest that in the rSC group, despite a weakened brain integration level (consistent with results in Figs. 2 and 4), the SCD remains positive. In cognitive neuroscience, previous studies have confirmed that the brain typically maintains a pro-integrated state during perceptual processing[75]. This indicates that although the rSC group's brain integration is reduced, it retains the ability to perform perceptual tasks (refer to Figure 1's RTT and Accuracy statistics, and Figure S2's ERP results).

However, in cases of cTBI with irreversible brain damage, the integration state exhibited during the perception process is significantly impaired, reflecting a more severe degree of brain damage. To more intuitively illustrate these individual differences, we calculated the SCD distributions for each group. A positive SCD value indicates that the trajectory lies above the balance line, while a negative value indicates it lies below (Figure 5C). The results clearly show that the HC group exhibited the strongest integration ability ($SCD_{HC} = 0.58$), the rSC group, though reduced ($SCD_{rSC} = 0.33$), still maintained positive values, whereas the SCD of the cTBI group fell within the negative range ($SCD_{cTBI} = -0.23$). These quantitative findings are fully consistent with the conclusions drawn from the earlier trajectory analysis.

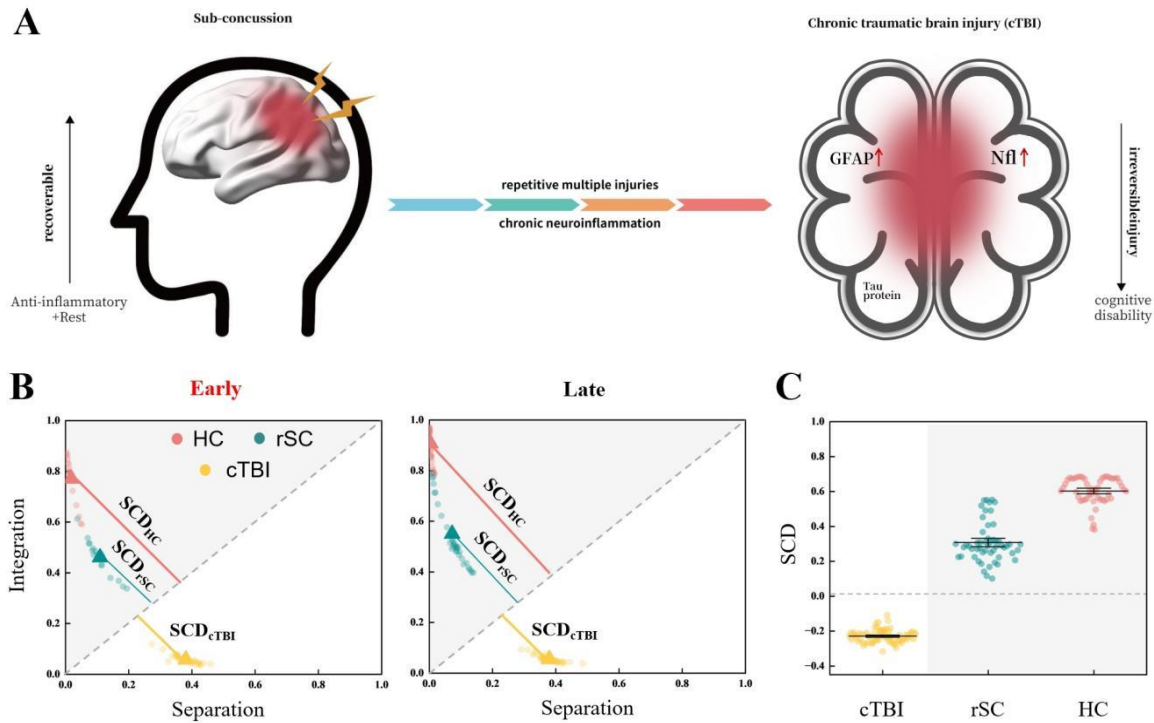

**Figure 5.** Quantitative Assessment of Cognitive Impairment in HC, rSC, and cTBI Groups Based on Separation-Integration Phase Trajectories. (A) Schematic diagram illustrating the deterioration pathway from subconcussion to cTBI. (B) Separation-integration phase diagrams of the three groups during the early and late periods, with separation-integration characteristic distance (SCD) values ($SCD_{HC}$, $SCD_{rSC}$,

SCD$_{cTBI}$) calculated for each group. (C) Statistical analysis of SCD from the phase diagrams: the HC group exhibited the largest mean distance (~0.58), the rSC group had a mean distance of ~0.33, and the cTBI group showed the smallest mean distance (~-0.23). A negative value ("-") indicates all data points in this group had greater separation values than integration values ($H_{se} > H_{in}$). Groups are color-coded: red (HC), blue (rSC), and yellow (cTBI). The dashed line denotes the ideal balance state where integration equals separation ($H_{in} = H_{se}$). The shaded area represents regions where integration > separation, while the area below the shaded region represents separation > integration.

By comparing the trajectory distributions between the early and late phases (Figure 5B and 2C), particularly for the HC and rSC groups, we observe that the overlap of distribution clusters in the early phase is significantly lower than in the late phase. This suggests that, compared to the late phase, the trajectories in the early phase may have a stronger ability to distinguish between HC and rSC. To validate this hypothesis, we constructed a machine learning classification model, using trajectory data from the early and late phases of the HC, rSC, and cTBI groups as model inputs, with disease categories as outputs. The validation results are presented in Figure 6.

*3.4. Early Integration-Segregation Trajectory in Predicting cTBI, rSC, and HC Group Differences*

This study utilized the separation-integration trajectory from both early and late phases as input features for four machine learning algorithms—Support Vector Machine (SVM), Naive Bayes (NB), K-nearest Neighbors (KNN), and Decision Tree (DT)—to predict brain injuries of varying severity levels, specifically in HC(n=24), rSC (n=21), and cTBI (n=25) groups (Figure 6A). During the training phase, the integration and separation values of the HC, rSC, and cTBI groups at each time point in the two states of consciousness were extracted as inputs, resulting in a total of 12 features.

The experimental results demonstrated that early-phase features exhibited exceptional algorithm robustness. All four machine learning algorithms achieved 100% classification accuracy under this feature set, with SVM and Naive Bayes algorithms showing near-perfect discriminative ability (AUC=0.971) (Figure 6B). In contrast, late-phase features displayed higher sensitivity to algorithm selection. While the SVM algorithm maintained 100% predictive accuracy (AUC=0.971), the performance of the other three algorithms decreased, with accuracy dropping to 93.3%. Comparative results indicated that the separation-integration trajectory features from the early phase had a stronger predictive ability for distinguishing between the HC, rSC, and cTBI. These features enabled all algorithms to achieve or approach 100% predictive precision. The distribution characteristics of the three conditions in Figure 5 can explain this result: the distribution positions of the three groups in the early phase were nearly completely isolated (Figure 5B and 5C), and this clear topological separation provided an ideal classification boundary for the machine learning models, reliably achieving the high-precision predictions shown in Figure 6.

However, compared to the early phase, the late-phase features showed significant differences, particularly in the overlapping distribution clusters between HC and rSC. When individual differences were considered, the degree of overlap would be further amplified (as shown in Figure 6, far exceeding the crossover state presented by the group average distribution), leading to blurred classification boundaries. This made the late-phase input features more sensitive to algorithm selection. Nevertheless, the SVM algorithm still maintained 100% predictive accuracy for late-phase features. This phenomenon stemmed from the inherent advantages of the SVM algorithm, including its stronger adaptability to small sample data and blurred classification boundaries[76,77]. It was precisely this characteristic that enabled SVM to effectively overcome the blurred boundary problem between HC and rSC distributions in the late phase, as shown in Figure 5.

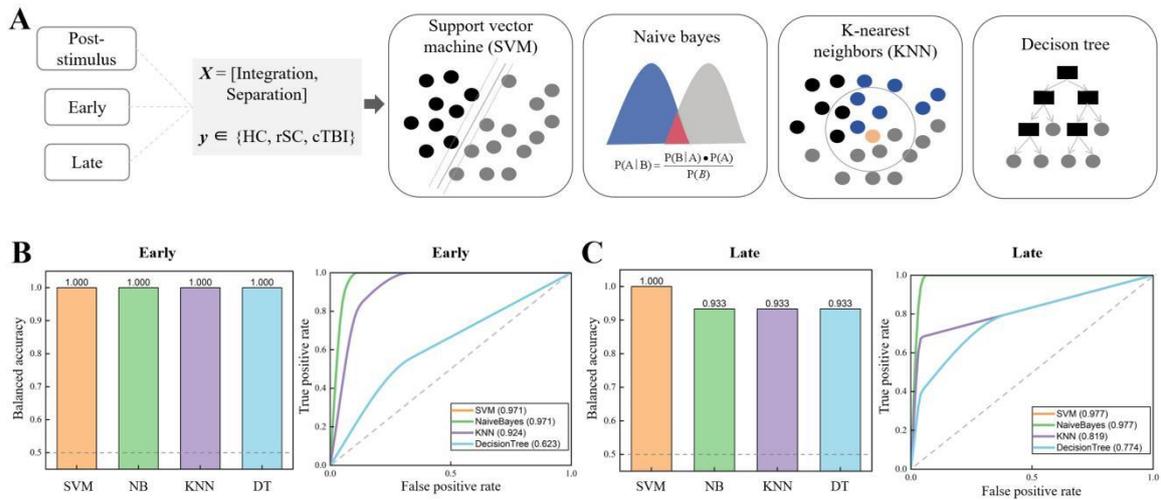

**Figure 6.** Predictive Accuracy Assessment of Early Integration-Separation Trajectory Patterns for Differentiating HC, rSC, and cTBI Groups Using Machine Learning Models. (A) Schematic representation of the machine learning framework, including the following algorithms: Support Vector Machine (SVM), Naive Bayes, K-Nearest Neighbors (KNN), and Decision Tree. (B) Comparative analysis of classification accuracy and receiver operating characteristic (ROC) curves (with area under the curve (AUC) values in parentheses) for integration-separation trajectory data in HC (n=24), rSC (n=21), and cTBI (n=25) groups at the early phase. (C) Comparative analysis of classification accuracy and ROC curves (with AUC values in parentheses) for integration-separation trajectory data in HC (n=24), rSC (n=21), and cTBI (n=25) groups at the late phase.

## 4. Discussion

Firstly, regarding the dynamic separation and integration patterns in HCs and rSCs, previous studies have focused primarily on long-term timescales of integration-segregation features to distinguish pathological from healthy states[21,78]. In contrast, the present study emphasizes millisecond-level dynamics of the integration-segregation balance. Our results reveal that HCs exhibit a significant peak in the whole-brain integration variable ($H_{in}$) at approximately 67 ms post-stimulation (Figure 2), whereas this peak amplitude is markedly attenuated in the rSC group. This finding emphasizes that, beyond long-term dynamic integration and segregation, millisecond-level short-term features also hold potential for predicting pathological conditions. As hypothesized, the early enhancement of integration observed at approximately 67 ms post-stimulation likely serves a critical role as a preparatory stage, optimizing and reorganizing neural connections to facilitate subsequent perceptual processing.

Furthermore, we focus on HCs and their early-phase pre-configuration patterns, specifically examining the key roles of the LTPC, OTPC, and PFPC in establishing the early preparatory state for subsequent perceptual processing. Our findings indicate that in the HC group, early-phase preliminary integration is accompanied by significantly more cross-regional connections within these cortical networks (Figure 4). Notably, previous studies have separately demonstrated that the LTPC, OTPC, and PFPC are critical for hippocampal relay function[65-67]. This aligns with our proposal that early-phase brain network configuration primarily functions to sustain hippocampal replay, serving as a preparatory stage for later perceptual processing. Consistent with generative network theory[28,29], the brain's perceptual processing is conceptualized as a generative process dependent on hippocampal replay[79]. However, prior research has lacked dynamic neural evidence supporting this process. In contrast, our comparative analysis of HC and rSC groups provides the first experimental evidence that HCs exhibit significant dynamic reconnection in LTPC, OTPC, and PFPC. This finding reaffirms the complexity of perceptual processing: It encompasses not only a straightforward feed-forward process but also a parallel mechanism involving self-organized pre-configuration (Figure 3).

In addition, prior studies have primarily focused on spatial deficits in specific brain regions associated with rSC[80,81], while overlooking temporal dynamic disruptions within these regions. Focusing on key cortical networks involved in hippocampal replay, the present study employs dynamic network analysis to reveal phase-specific attenuations in EEG functional connectivity among rSC patients.

Our findings indicate that during the early phase (~0–100 ms), cross-regional connectivity is significantly reduced in the rSC group. Notably, within the 220–270 ms window, connectivity in the LTPC, OTPC, and PFPC is weakened in rSC patients (Figure 4B); however, these connections exhibit a delayed compensatory rebound, which may serve to complete the preparatory stage for subsequent perceptual processing. This observation underscores the temporal precision of functional impairments in brain regions linked to clinical concussion symptoms, providing richer dynamic insights to deepen our understanding of neural deficits in concussion patients[82].

Finally, we discuss the predictive power of early-phase separation-integration features for distinguishing between HC, rSC, and cTBI groups. Previous studies have explored the dynamic balance between integration and separation, often quantifying this balance using the arithmetic difference between integration and separation metrics[21]. In contrast, the present study introduces a geometric approach to characterize the balance between dynamic integration and separation. Specifically, we developed a method to quantify traumatic brain injury-related abnormalities by measuring the signed centroid distance (SCD)—defined as the distance between the centroid of separation-integration phase trajectories and the equilibrium line ($H_{se} = H_{in}$)—from a dynamic network perspective. Our findings reveal that during the early phase, rSCs group exhibit significantly lower SCD values compared to HCs, while cTBIs show even smaller absolute SCD values, with their phase trajectories distributed below the equilibrium line ($H_{se} = H_{in}$), corresponding to negative distances (Figure 5). These results indicate that signed SCD not only discriminates the severity of recoverable rSC pathology but also identifies cTBI-related abnormalities via its negative value characteristics. Validated by machine learning analyses, early-phase integration-separation metrics achieved robust differentiation between HC, rSC, and cTBI groups, highlighting the diagnostic potential of early neural activity features for detecting subtle brain function abnormalities.

Nevertheless, this study has several limitations that merit consideration, particularly regarding the exploration of mechanisms underlying early pre-configuration of brain perceptual processing.

First, the rSC group was restricted to young male participants due to the nature of parachute tasks, which constrains the generalizability of our findings. Age and sex are well-established modifiers of EEG dynamics: advancing age, for instance, correlates with marked declines in sensory perception efficiency[83,84], potentially exerting significant effect on perceptual pre-configuration. To address this, future studies should recruit more diverse samples—encompassing broader age ranges and both sexes among healthy controls and subconcussive populations—to clarify how age- and sex-related factors modulate the dynamics of brain pre-configuration. Second, we defined the early processing window as 0–200 ms, with the initial 100 ms identified as critical for pre-configuration dynamics (Figure 2 and 4), consistent with prior reports[37]. The 100–200 ms interval appears to serve a rebound function sustaining subsequent information processing in healthy individuals; however, the predefined 200 ms endpoint remains debated. This uncertainty may stem from constraints of our experimental paradigm, specifically the use of a GO-NOGO task with 50 ms transient visual stimuli. Future work should systematically evaluate the variability and precision of this temporal range to refine its biological validity. Third, our analysis focused on activated regions including the LTPC, OTPC, and PFPC—core regions in generative models of memory retrieval. While we highlighted the critical role of hippocampal replay in memory retrieval, alongside contributions from the medial temporal lobe (e.g., entorhinal cortex[85]) and the aforementioned cortices, these structures span different brain layers, and single-modality techniques (e.g., scalp EEG) limit comprehensive characterization of their interactions. To address this, multi-modal approaches integrating intracranial EEG and fMRI—offering high spatiotemporal resolution—should be employed to enable a holistic understanding of how all relevant brain regions contribute to pre-configuration dynamics.

## 5. Conclusion

In the brain's bottom-up processing hierarchy, cortical neural activity critically regulates perceptual information transmission. While classical theories highlight steady-state oscillations (e.g., gamma traveling waves in primary sensory areas) as foundational to sensory responses[86,87], our study demonstrates that the brain—operating as a self-organized complex system—dynamically reorganizes its functional architecture during early pre-configuration stage (0–200 ms) to optimize subsequent conscious processing. Using separation-integration dynamics, we show that HCs exhibit a transient whole-brain integration peak ($H_{in}$) at ~67 ms post-stimulus (Figure 2A), a preparatory state that supports later cognitive integration (200–600 ms). In contrast, individuals with rSC display attenuated $H_{in}$ (Figure 2D) and delayed compensatory integration (220–270 ms; Figure 4), whereas those with cTBI show irreversible network segregation ($H_{in} < H_{se}$; Figure 5B). These findings validate the signed centroid distance (SCD) as a robust metric for quantifying dysfunction severity across healthy, rSC, and cTBI states (Figure 5C), providing a novel framework to probe dynamic neural pathology in traumatic brain injuries. Specifically, by combining machine learning with analyses of integration-separation dynamics across early and late phases, our findings indicate that early-phase features exhibit superior performance in predicting the pathological state of rSC when distinguishing it from HC and cTBI. Collectively, our results reveal that enhanced early-phase neural integration prepares the brain for subsequent fine-grained information processing in awake states, offering new insights into the dynamic evolution of neural computational mechanisms across traumatic brain injury subtypes.

**CRediT authorship contribution statement**

**Jiajia Li:** Writing–original draft, Writing–review & editing, Conceptualization, Investigation, Visualization, Analysis; **Zhenzhen Yu:** Writing–original draft, Conceptualization, Visualization, Analysis; **Zhenghao Fu:** Investigation; **Guozheng Xu:** Writing–review & editing, Investigation, Supervision; **Jian Song:** Writing–review & editing, Investigation, Supervision.

**Declaration of Competing Interest**

Authors declare that they have no competing interests.

**Acknowledgments**

We thank the Department of Neurosurgery of the General Hospital of Central Theater Command for their medical support, as well as all participants for their contributions.

**Funding**

This work was supported by the National Natural Science Foundation of China (Grant Nos. 12572068, 81870863).

**Data and materials availability**

The dataset for this study will be made available upon request. The Chronic Traumatic Brain Injury EEG dataset is available from OpenNEURO[43].

# Supplementary Material for
# Early Preconfiguration Failure：A Novel Predictor of the Repetitive Subconcussion

Jiajia Li [1,2], Zhenzhen Yu [1], Zhenghao Fu [2,3,4], Guozheng Xu [2,3,*], Jian Song [2,3,*]

[1]College of Information and Control Engineering, Xi'an University of Architecture and Technology, Shaanxi, Xi'an, 710055, China.

[2]Department of Neurosurgery, General Hospital of Central Theater Command, Wuhan, 430070, China.

[3]The First School of Clinical Medicine, Southern Medical University, Guangzhou 510515, China.

[4]Department of Neurosurgery, Guangdong Sanjiu Brain Hospital, Guangzhou 510510, China.

*Corresponding authors at: Department of Neurosurgery, General Hospital of Central Theater Command, 627 Wuluo Road, Wuhan 430070, China.

**Email addresses:** songjian0505@smu.edu.cn (J. Song), xu-gz@163.com (G. Xu).

**This PDF file includes:**

    Figures S1 to S2
    Table S1

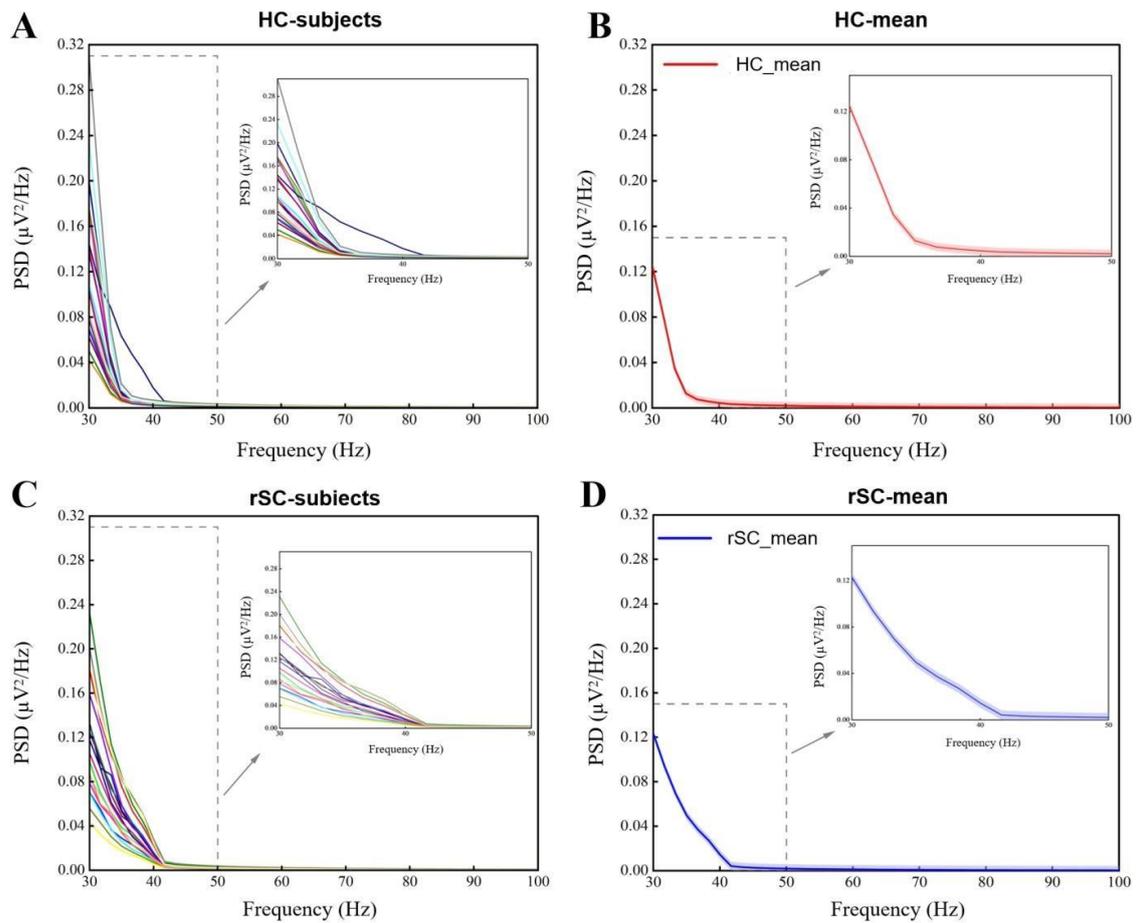

**Figure S1.** Spectral power in the gamma band (30–100 Hz) for EEG responses. (A) Spectrogram of each subject in the HC group. (B) Mean spectrogram of all subjects in HC group ± mean standard error (SEM). (C) Spectrogram of each subject in the SC group. (D) Mean spectrogram of all subjects in rSC group ± mean standard error (SEM). The gamma oscillation rhythm is significantly stronger in the 30–40 Hz band. The red line indicates the HC group mean; the blue line indicates the rSC group mean.

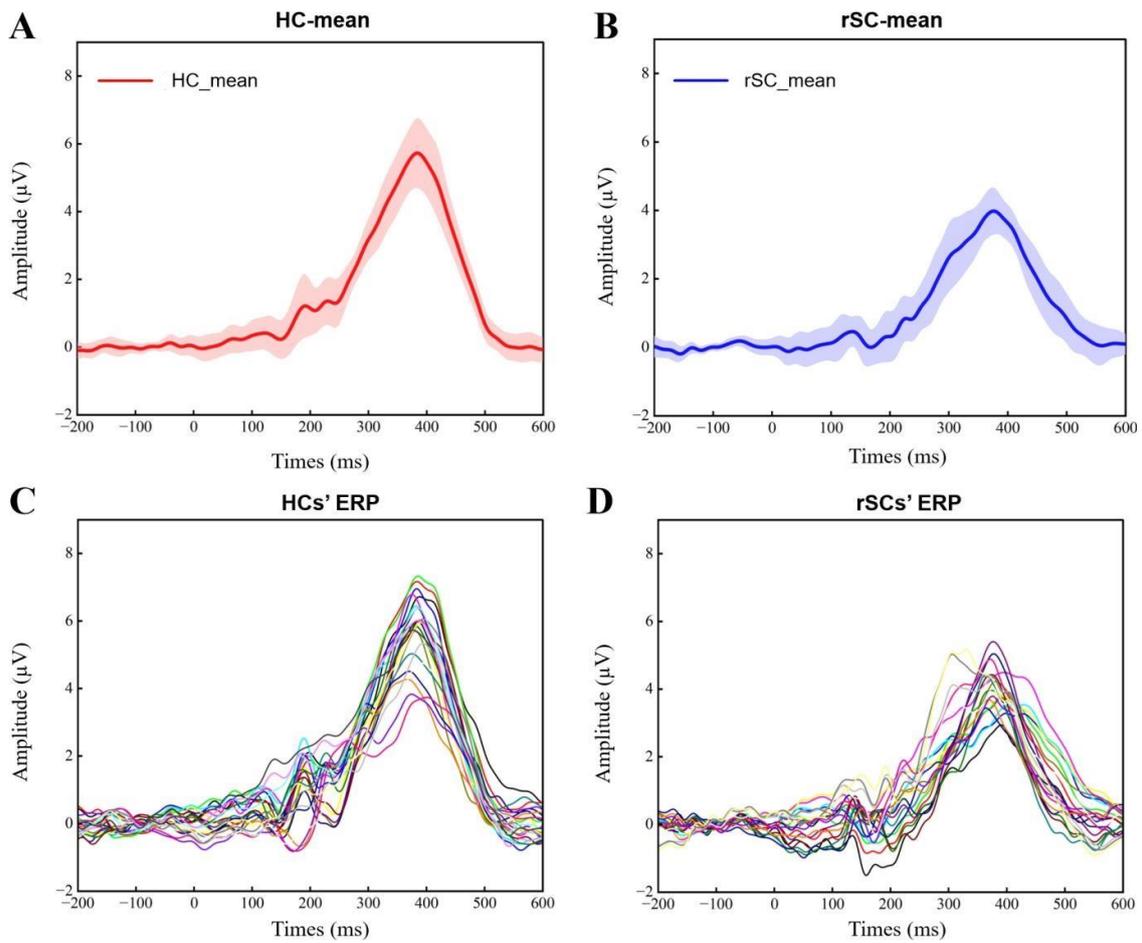

**Figure S2.** Event Related Potentials(ERPs) for HC and SC groups. (A) Averaged ERP for the HC group ± mean standard error (SEM). (B) Averaged ERP for the rSC group ± mean standard error (SEM). (C) ERPs of individual subjects in the HCs. (D) ERPs of individual subjects in the rSCs. The red line indicates the HC group mean; the blue line indicates the rSC group mean.

**Table S1.** Number of trials per participant in HC and rSC.

| HC Subject ID | Trials (per participant) | rSC Subject ID | Trials (per participant) |
|---|---|---|---|
| HC_01 | 283 | rSC_01 | 230 |
| HC_02 | 299 | rSC_02 | 245 |
| HC_03 | 297 | rSC_03 | 287 |
| HC_04 | 298 | rSC_04 | 277 |
| HC_05 | 298 | rSC_05 | 284 |
| HC_06 | 296 | rSC_06 | 281 |
| HC_07 | 295 | rSC_07 | 268 |
| HC_08 | 265 | rSC_08 | 277 |
| HC_09 | 278 | rSC_09 | 226 |
| HC_10 | 298 | rSC_10 | 230 |
| HC_11 | 299 | rSC_11 | 224 |
| HC_12 | 297 | rSC_12 | 187 |
| HC_13 | 284 | rSC_13 | 256 |
| HC_14 | 291 | rSC_14 | 261 |
| HC_15 | 298 | rSC_15 | 286 |
| HC_16 | 281 | rSC_16 | 300 |
| HC_17 | 298 | rSC_17 | 300 |
| HC_18 | 294 | rSC_18 | 260 |
| HC_19 | 300 | rSC_19 | 296 |
| HC_20 | 275 | rSC_20 | 300 |
| HC_21 | 294 | rSC_21 | 296 |
| HC_22 | 296 | | |
| HC_23 | 292 | | |
| HC_24 | 297 | | |
| Mean ± SD | 291.792 ± 9.198 | Mean ± SD | 265.286 ± 31.379 |